\documentclass[11pt]{article}
\begin{document}

\title {New Results on the Parameterisation of  Complex Hadamard Matrices }

\author{P Di\c t\u a \\Institute of Physics and Nuclear Engineering,\\
P.O. Box MG6, Bucharest, Romania\\
email: dita@zeus.theory.nipne.ro}

\maketitle

\begin{abstract}

In this paper we provide an analytical procedure which leads to a system of $(n-2)^2$ polynomial equations whose solutions  give  the  parameterisation of the complex $n\times n$  Hadamard matrices. It is shown that in general the Hadamard matrices depend on a number of arbitrary phases and a lower bound for this number is given.  The moduli equations define interesting geometrical objects whose study will shed light  on the  parameterisation of Hadamard matrices, as well as  on some interesting geometrical varieties defined by them.

\end{abstract}

\section{Introduction}

Quantum information theory whose main source comes of  a few astonishing features  in the foundations of quantum mechanics is the theory of that kind of information which is carried by quantum systems from the preparation device to the measuring apparatus in a quantum mechanical experiment, see e.g. \cite{We}.
Defining new concepts like entangled states, teleportation or dense coding one hopes to  be able to design and construct new devices, like quantum computers, which will be useful in solving many ``unresolvable'' problems by the classical methods. Recently the mathematical structure which is behind such miracle machines was better  understood by establishing a one-to-one correspondence between quantum teleportation schemes, dense coding schemes, orthogonal bases of maximally entangled vectors, bases of unitary operators and unitary depolarizers by showing that given any object of any one of the above types one can construct any object of each of these types by using a precise procedure. See 
Vollbrecht and Werner \cite{VW} and Werner \cite{We1} for details. The construction procedure will be efficient  to the extent that the unitary bases can be generated, and the construction of these  bases makes explicit  use of the  complex Hadamard matrices and  Latin squares. The aim of this paper is to provide a procedure for the parametrisation of the complex Hadamard matrices for
an  arbitrary integer $n$. More precisely we will obtain a set of $(n-2)^2$ equations whose solutions will give all the complex Hadamard matrices of size $n$.
Complex $n$-dimensional Hadamard matrices are unitary $n\times n$ matrices 
whose entries have modulus $1/\sqrt{n}$. 

The term {\it Hadamard matrix} has its root in the Hadamard's paper \cite{Ha}, where he gave the solution   to the question of the maximum possible absolute  value of the determinant of a complex $n\times n$ matrix whose  entries are bounded by some constant, which, without loss of generality, can be taken equal to unity. Hadamard has shown that the maximum is attained by complex unitary matrices whose  entries have the same modulus  and he asked the question if the maximum can also  be attained by orthogonal matrices. These last matrices have come to be known as {\it Hadamard matrices} in his honor, and have many applications in  combinatorics, coding theory, orthogonal designs, quantum information theory, etc., and a good reference about the obtained results is Agaian \cite{Ag}. 

However the first complex Hadamard matrices were found by Sylvester \cite{Sy}. He observed that if  $a_i,\,\, i=0,1,\dots,n-1$ denote the  solutions of the equation $x^n-1=0$ for a prime $n$ then the Vandermonde matrix
\[
{1\over\sqrt{n}}\left(\begin{array}{ccccc}
1&1&1&\cdots&1\\
1&a_1&a_1^2&\cdots&a_1^{n-1}\\
\cdots&\cdots&\cdots&\cdots&\cdots\\
1&a_{n-1}&a_{n-1}^2&\cdots&a_{n-1}^{n-1}
\end{array}
\right)
\]
is unitary and Hadamard. In the same paper Sylvester found a method to obtain a Hadamard matrix of size $m n$ if one knows two Hadamard matrices of order $m$ and respectively $n$ by taking their Kronecker product.
Soon after the publication of the paper by Hadamard the interest was mainly on the {\it real} Hadamard matrices such that the Sylvester contribution fell into oblivion and the  {\it complex Hadamard matrices} have been much later  reinvented  
in a particular case: only those matrices whose entries are $\pm\, 1,\pm\, i$ where $i=\sqrt{-1}$. 

Nevertheless a few other problems apparently unrelated to complex Hadamard matrices were those connected with bounds on polynomial coefficients when the indeterminate runs on the unit circle. They are better expressed in terms of the  discrete Fourier transform. For any finite sequence $x=(x_0,x_1,\dots,x_{n-1})$ of $n$ complex numbers, its (discrete) Fourier transform is defined by
$$y_j=n^{-1/2}\sum_{k=0}^{n-1}\,x_k\,e^{2\,i\,\pi\,kj/n}\quad  j=0,1,\dots,n-1$$If the components $x_k,y_k$ are such that $|x_k|=|y_k|=1$ for $k=0,1,\dots,n-1$ the sequence $x$ is called bi-unimodular. The existence of a bi-unimodular sequence of side $n$ is equivalent to the existence of a complex circulant Hadamard matrix of side $n$; a circulant matrix is obtained by circulating its first row, in our case  the components of the  vector $x/\sqrt{n}$. Now the Gauss sequence
$$
x_k=\left\{\begin{array}{ll}e^{2\,i\,\pi(ak^2+bk)/n},\,\, a,\,b\in {\mathbf{Z}},\, a\,{\rm coprime\,to}\,\, n,\, k=0,1,\dots,n-1 \, & \mbox{for $n$  odd}\\
e^{k^2\,i\,\pi/n},\,\qquad  k=0,1,\dots,n-1& \mbox{for $n$  even}
\end{array}\right.
$$
is a bi-unimodular sequence \cite{BS}. The problem of the complete determination  of all bi-unimodular sequences is still open, despite the problem is simpler than the parameterisation of arbitrary complex Hadamard matrices. However this approach gave the first non-trivial examples of  complex Hadamard matrices for $n\ge 6$.

A  step towards its  solution  was the reduction of the bi-unimodular problem to the problem of finding all cyclic $n$-roots  \cite{Bj},  that are given by the following  system of equations over $\mathbf{C}$
\begin{eqnarray}
\left\{\begin{array}{r}
z_0+z_1+\cdots +z_{n-1}=0,\\
z_0z_1+z_1z_2+\cdots +z_{n-1}z_0=0,\\
z_0z_1z_2+z_1z_2z_3+ \cdots + z_{n-1}z_0z_1=0,\\
\cdots\cdots\cdots\\
z_0z_1\cdots z_{n-1}=1
\end{array}
\right.\label{sys}
\end{eqnarray}
Note that the sums are cyclic and contain just $n$ terms and are not the elementary symmetric functions for $n\ge 4$.
The relation between $x$ and $z$ is $z_j=x_{j+1}/x_j$. All cyclic $n$-roots have been found for $2\le n \le 8$; see Bj\"orck and Fr\"oberg \cite{BF,BF1}. The formalism we will develop in the paper is more general showing that the parameterisation of complex Hadamard matrices is more complicated than the finding of all cyclic $n$-roots of the sytem (\ref{sys}).  Using our approach we find,  e.g. when $n=6$, the following matrix which is not contained in the above solutions
$${1\over\sqrt{6}}\left(
\begin{array}{rrcccc}
1&1&1&1&1&1\\
1&-1&i&-i&-i&i\\
1&i&-1&e^{it}&-e^{it}&-i\\
1&-i&-e^{-it}&-1&i&e^{-it}\\
1&-i&e^{-it}&i&-1&-e^{-it}\\
1&i&-i&-e^{it}&e^{it}&-1
\end{array}\right)$$
matrix that depends on an arbitrary phase.

The parameterisation of complex Hadamard matrices is a special case of a more general problem: that of reconstructing the phases of a unitary matrix from
the knowledge of the moduli of its entries, problem which  was a fashionable one at the end of
eighties of the last century in the high energy physics community, see Auberson\cite{Au}, Bj\"orken and Dunietz \cite{BD}, Branco and Lavoura \cite{BL}, Auberson {\em et al.} \cite{AMM}. An existence
theorem as well as an estimation  for the number  of solutions
was obtained by us \cite{Di2}.  The particle  physicists abandoned the problem when they realised that for $n \ge 4$ there exists a continuum of solutions, i.e. solutions depending on arbitrary phases, result that was considered uninteresting from the physical point of view. In our opinion, the reason was the difficulty of the problem; since the experiments provide only the squares of the moduli, the first problem is to decide if from the experimental results, which in the best case generate  a doubly stochastic matrix, one can reconstruct a unitary matrix, or a unistochastic matrix. Only for $n=3$ there exists a unambigous
procedure. For $n\ge 4$ there are no known necessary and sufficient conditions to separate the unistochastic matrices from the doubly stochastic ones \cite{Zy}.

Almost in the same time  the complex Hadamard matrices came out  in the construction  of some $*$-subalgebras in finite von Neumann algebras, see Popa \cite{Po}, de la Harpe and Jones \cite{HJ} and  Munemasa and Watatani\cite{MW} . In the last two papers one construct complex Hadamard matrices not of Sylvester type when $n$ is a prime number such that $n\equiv\pm 1$ (mod 4). A little later  Haagerup \cite{Haa}  obtained the first example of a 6-dimensional matrix which is not a solution of the  system of equations (\ref{sys}).

In this paper we make use  of a few analytic techniques from the operator contraction theory and the factorization of unitary matrices to obtain a convenient reprezentation of unitary matrices of arbitrary order $n$ that leads us easily to a system of $(n-2)^2$   trigonometric (or equivalently polynomial) equations whose solutions give all the complex Hadamard matrices of order $n$. Our approach is also useful for finding {\it real } Hadamard matrices, being complementary to the combinatorial approach almost exclusively used until now.

The paper is organized as follows: in Section 2 the equivalence of the complex Hadamard matrices is reviewed. In Section 3 a theorem  showing the existence of the complex  Hadamard matrices for every integer $n$ is stated and an upper bound on the number of continuum solutions is obtained. Section 4 contains an one-to-one parametrisation of unitary matrices written as block matrices and in the next Section an application of the obtained formulae is given. In Section 6 an other parameterisation of unitary matrices is given under the form of a product of $n$ diagonal phase matrices interlaced with $n-1$ orthogonal matrices each one generated by a real vector from ${\bf R}^n$. This form is convenient because it leads to a simpler form for the moduli equations and in the same time  we consider  it more appropriate for designing software packages for solving these equations. In Section 7 we show how to derive the moduli equations as trigonometric equations and give a few particular solutions for $n=6$. In Section 8 the problem is reformulated as an algebraic geometry problem and we show that the parameterisation of Hadamard matrices can produce interesting examples for many problems currently under study in this field.  The paper ends with Conclusions.

\section{Equivalence of complex Hadamard matrices}

Complex $n$-dimensional Hadamard matrices being  unitary  matrices 
whose entries have modulus $1/\sqrt{n}$, the natural class 
of looking  for complex Hadamard matrices is the unitary group
${U}(n)$.

 The unitary group ${U}(n)$ is the group of automorphisms of
the Hilbert space $({\mathbf C}^{n}, (\cdot,\cdot))$ where $(\cdot,\cdot)$ denotes the
Hermitian scalar  product $(x,y)=\sum_{i=1}^{i=n}\,\overline{x_i}\,y_i$ and  the bar denotes the complex conjugation. If
$A_n\in {U}(n)$ by $A_n^*$ we  denote the adjoint matrix and unitarity implies
$A^*_n\,A_n=A_n\,A^*_n=I_n$. It  follows
that $det\, A_n= e^{i\,\varphi}$, where $\varphi$ is a phase, and
$dim_{\bf R}\,{U}(n)=n^2$.

Because in any group the product of two arbitrary elements is again an element
of the  group there is a freedom in choosing the "building" blocks to be used
in a definite application. In the case of a complex Hadamard matrix the multiplication of a row
and/or a column by an arbitrary phase factor does not change its properties and
consequently we can remove the phases of a row and column taken arbitrarily.
Taking into account that property we can write
\begin{eqnarray}
A_n=d_n\,\tilde{A_n}\, d_{n-1}\end{eqnarray}
where $\tilde{A_n}$ is a matrix with  all the elements of the first row and
of the first column positive numbers and 
 $d_n=(e^{i\varphi_1},\dots,e^{i\varphi_n})$  and
$d_{n-1}=(1,e^{i\varphi_{n+1}},\dots,e^{i\varphi_{2n-1}})$
are two diagonal phase matrices. In the following we will consider that $A_n \equiv
 \tilde{A_n}$, i.e.  $A_n$ will be a matrix with positive entries in the first row and the first  column.

Since a unitary matrix is parameterised by $n(n-1)/2$ angles and $n(n+1)/2$
phases \cite{Di} the above equivalence relation tell us that the number of remaining phases is 
$n(n+1)/2-(2n-1)=(n-1)(n-2)/2$, and so the  number of free real parameters entering a
unitary matrix is reduced from $n^2$ to $n^2-(2n-1)=(n-1)^2$.

Secondly we can permute any  rows and/or  columns and get an
equivalent unitary matrix. This procedure can be seen as a multiplication of $A_n$ at
left and/or right by an arbitrary finite number of the simplest permutation unitary
matrices $P_{ij},\,\,   i\neq j,\,\, i,j=1,\dots,n$, whose all diagonal entries
but $a_{ii}$ and $a_{jj}$ are equal to unity, $a_{ii}=a_{jj}=0,\,\,
a_{ij}=a_{ji}=1,\,\, i\neq j$ and all the other entries vanish. Both the diagonal phase and permutation matrices generate subgroups of the unitary ${U}(n)$ group; so we may consider them as gauge subgroups, i.e. any element of ${U}(n)$ is defined modulo the action of a finite number of the above transformation, which has as consequence  a standard representation for unitary matrices. We consider that the group generated by the above two subgroups deseves to be independently studied since its orbit structure could shed light on many important issues from information theory and stochastic matrices.

The above two equivalence conditions are   those found by Sylvester \cite{Sy} for the Hadamard matrices, but  in fact they are valid for ${U}(n)$ which is   invariant with respect to   the product of an arbitrary number of the above transformations.

Besides for Hadamard matrices we will not distinguish between $A_n$ and its complex conjugated matrix 
$\bar{A}_n$, the complex conjugation  being equivalent to the sign change of all phases $\varphi_i \rightarrow - \varphi_i$ entering the parametrisation.
 More generally  we shall consider equivalent two matrices whose phases can be obtained each other by an arbitrary non-singular linear transformation with constant rational coefficients. As we will see later the complex Hadamard matrices  depend in general on a number of arbitrary phases  and the above condition says that we will consider only the most general form of the solution  and not those particular forms obtained by prescribing definite values to the (arbitrary) phases entering the parameterisation. In this sense we can say that there is only one complex Hadamard matrix of order $4$, that found by Hadamard \cite{Ha}, all the others, including those with all entries real numbers, being  particular cases of the complex one. Other authors speak in this case of   non-equivalent or a continuum   of solutions \cite{Haa}.
 We consider
that the above conditions are the only a priori  equivalence criteria we can
impose on Hadamard matrices, i.e. will consider equivalent any two matrices that can be made equal by applying  them a finite number of the above transformations.

\section{Existence of complex Hadamard matrices}

The parameterisation of a unitary matrix by the moduli of
its entries is very appealing, and in the case of Hadamard matrices compulsory,
although it is not a natural one in the general case. A natural parameterisation would be
one whose parameters are free, i.e. there are no supplementary restrictions
upon them to enforce unitarity. In this sense natural parameterizations are the
Euler-type parameterisation by Murnagham \cite{Mu}, or that found by us
\cite{Di}.

The problem we rose in \cite{Di2} was to what extent the knowledge of the
moduli $|a_{ij}|$ of an $n\times n$ unitary matrix $A_n=(a_{ij})$ determines
$A_n$. Implicitly we supposed that $A_n$ is  parameterized by $n^2$ independent
parameters. But from what we said before we know that we may
ignore $2n-1$ phases entering the first row and the first column and consequently the
number of independent parameters reduces to $(n-1)^2$, that coincides with the
number of independent moduli implied by unitarity. If we identify the
parameters to the moduli they will be lying within the simple domain
$$D=(0,1)\times\dots\times (0,1)\equiv (0,1)^{(n-1)^2}$$
where the above notation means that the number of factors entering the
topological product is $(n-1)^2$. We excluded only the extremities of each
  interval, i.e. the points  $0$ and $1$ that is a zero measure set whitin
${U}(n)$ and has no relevance to the parameterisation of complex Hadamard
matrices.

Thus, in principle, we can parameterise an $n\times n$ unitary rephasing
invariant matrix by the upper left corner moduli; we exclude the moduli of the
last row and of the last column since they follow from  unitarity. Nothing remains but to check if the new parameterisation is
one-to-one. A solution to the last problem is the following: start with a
one-to-one parameterisation of ${U}(n)$ and then change the coordinates taking as
new coordinates the moduli of the $(n-1)^2$ upper left corner entries (and
$2n-1$ ignorable phases). Afterwards use the implicit function theorem to find
the points where the new parameterisation fails to be one-to-one. The
corresponding variety upon which the application is not a bijective one is
given by setting to zero the Jacobian of the transformation. One gets that generically for
$n\ge 4$ the unitary group ${U}(n)$ cannot be fully parametrised by the moduli
of its entries, i.e. for a given set of moduli there could exist a continuum of  solutions,  but this negative result is good for the parameterisation of 
Hadamard matrices by decreasing the number of independent solutions taking into account the equivalence conditions discussed in the previous section.

If the moduli are outside  of the above variety an upper bound for the multiplicity is
$2^{n(n-3)\over2}$. However in the case of Hadamard matrices the equivalence
constraints  reduce this number to lower
values than the above upper bound. The bound is saturated for $n=3$ when there
is essentially  only one complex  matrix, i.e. for given  moduli values for the first row and column entries compatible with unitarity, the sole freedom is an arbitrary phase. 
If we denote the relevant squared moduli by $m_1, m_2, m_3, m_4$ and the phase by $\varphi$ then the compatibility condition has the form
\begin{center}
$
-1\leq \cos\varphi=(-1+2m_1-m_1^2+m_2+m_3+m_4-m_1m_2-m_1 m_3-m_2 m_3 -$
 \vskip2mm
$2 m_1 m_4 - m_1 m_2 m_3m_1^2 m_4)/2\sqrt{m_1 m_2 m_3 (1-m_1-m_2)(1-m_1-m_3)} \leq 1$
 \end{center}
This is also the necessary and sufficient condition which the squared moduli 
 $m_i,\,\,i=1,\dots,4$, have to satisfy in order to
 obtain a unistochastic  matrix from a general  doubly stochastic matrix.
Because unitary matrices of arbitrary dimension do exist and on the
other hand the number of independent essential parameters of a ${U}(n)$ matrix
is $(n-1)^2$   the following is true:

\newtheorem{so}{Theorem}
\begin{so}
Suppose $(x_1,\dots,x_{n^2})$ is  a co-ordinate system on the unitary
group ${U}(n)$  consisting of $n(n-1)/2$ angles each one taking
values in $[0,\pi/2]$ and $n(n+1)/2$ phases taking values in $[0,2\pi)$. By
discarding $2n-1$ non-essential phases the number of co-ordinates reduces to
$(n-1)^2$, $( x_1,\dots,x_{(n-1)^2})$, that coincides with the number of
independent moduli $(m_1,\dots,m_{(n-1)^2})$ implied by unitarity. Taking as
new co-ordinates the moduli $m_i,\,\, i=1,\dots,(n-1)^2,$ the new
parameterisation is generically not one-to-one for $n \geq 4$,  the non-uniqueness
variety being  obtained by setting to zero the Jacobian of the transformation
\begin{eqnarray}
{\partial(m_1,\dots,m_{(n-1)^2})\over\partial(x_1,\dots,x_{(n-1)^2})}=0\label{jac}
\end{eqnarray}
Outside this variety the number of discrete  solutions $N_s$ satisfies $1\leq
N_s\leq 2^{{n(n-3)\over 2}}$ and  on the variety described by (\ref{jac}) there is a continuum
of solutions. In the special case of complex Hadamard matrices all the
solutions are given by the system of  trigonometric equations 
\begin{eqnarray} 
  m_i^2(x_1,\dots,x_{(n-1)^2})={1\over{n}}\,,\qquad
i=1,\dots,(n-1)^2\label{mod}
\end{eqnarray}
Suppose  we know the irreducible components of  the variety (\ref{jac})
and let $r(n)$ be the rank of the system (\ref{mod}) in every irreducible
component,  then every solution of (\ref{mod})  in such an irreducible component  will depend upon  $(n-1)^2-r(n)$
arbitrary parameters and the number of (continuum) solutions satisfies $1\leq
N_s\leq 2^{r(n)-1-n(n-1)/2}$.
 \end{so}

{\em Proof.} In the general case Eqs.(\ref{mod}) have the form
\begin{eqnarray}
m_i^2(x_1,\dots,x_{(n-1)}^2)=a_i,\,\,\,\,\,{\rm where}\,\,a_i\in (0,1)\,,\quad i=1,\dots,(n-1)^2\label{mod1}
\end{eqnarray}
The parameters $a_i$ generate a doubly stochastic matrix. The Eqs.(\ref{mod1}),  as we will see later, are trigonometric equations in our parameterisation, and consequently the multiplicity of the solutions may arise from the
two possible phase solutions
 for all values of sine or cosine functions that
satisfy (\ref{mod1}).  The number of independent phases is $(n-1)(n-2)/2$
and  taking  into account   that we consider $A_n$ and $\bar{A_n}$ as
equivalent matrices, condition which halves the number of solutions, the above
bound for $N_s$ follows. A similar argument establishes the upper bound for the number of continuum solutions.

 For $n=3$ the Jacobian is positive and  $
1\leq  N_s\leq 1$, which
implies the existence of one complex matrix irrespective of the values $a_i$,
compatible with unitarity.

 It is easily seen that the equations which correspond to the first row and the first column entries have a unique solution and the number of equations reduces to $(n-2)^2$. Indeed, 
because these entries  are positive we can take the following parameterisation in terms of $2\,n-3 $ angles, e.g. for the first row
\[
(a_{11},\dots,a_{1n})=(cos\,\chi_1,sin\,\chi_1\,cos\,\chi_2,\dots,sin\,\chi_1\dots sin\,\chi_{n-1})
\]
and similarly for the first column.
The Eqs.(\ref{mod1}) give the unique solution
$$cos^2\,\chi_k=\frac{a_k}{\prod_{i=1}^{k-1}(1-a_i)},\quad k=1,2,\dots,n-1$$ 
where $a_k=|a_{1k}|^2, \,  k=1,2,\dots,n-1$.
In the case of Hadamard matrices one gets

$$cos\,\chi_k=\frac{1}{\sqrt{n+1-k}},\quad k=1,2,\dots,n-1$$ 
and the same solution for the angles parameterising the first column. In this way the number of equations reduces to $(n-1)^2-(2\,n-3)=(n-2)^2$ and the upper bound for the continuous solutions may be written as $1\leq N_s\leq 2^{r(n)-1-(n-2)(n-3)/2}$, where $r(n) $ is the rank of the reduced system. Even so the number of equations grows quadratically with $n$ which shows that even for moderate values of $n$ the problem is not easy to solve.

In conclusion we have  a system of trigonometric equations whose solutions will give all the complex Hadamard matrices, but to get  effective we have to start with a one-to-one parameterisation of unitary matrices in order  to find the explicit form of the $(n-2)^2$ equations and try to solve them. In the following Section we will provide one of the two  parameterisations of unitary matrices that we will use in the paper.

\section{Parameterisation of unitary matrices}

The aim of this section is to provide  a one-to-one parameterisation
 of unitary matrices that will be   useful in  describing the  complex Hadamard matrices. We shall present two such parameterisations and  for the the first one
we follow closely our paper \cite{Di}  showing  here only the  points which are important in the following. The
algorithm we provide is a recursive one, allowing the parameterisation of
$n\times n$ unitary matrices through the parameterisation of lower dimensional
ones. The parameterisation will be one-to-one and given in terms of $a(n)$
angles taking values in $[0,\pi/2]$ and $\varphi(n)$ phases taking values in
$[0,2\pi)$ such that the application  $$A_n(A_n\in {{U}}(n),
A_nA_n^*=I_n)\rightarrow E=(0,\pi/2)^{a(n)}[0,2\pi)^{\varphi(n)}\subset
{\mathbf{ R}}^{n^2}$$  is bijective. Always in the following  the ends
of the interval $[0,\pi/2]$ will be obtained by continuation in the relevant parameters, if necessary.

  The starting point is the
partitioning of the matrix $A_n\in{ U}(n)$ in blocks
\begin{eqnarray} 
A_n=\left(
\begin{array}{cc}A &B\\ C&D\end{array}\right)
 \label{block}
\end{eqnarray}
 For definiteness we
suppose the order  of $A$  is equal to  $m$  with $m\leq n/2$. The blocks entering (\ref{block})
are  contractions as follows from unitarity 
\begin{eqnarray}
A\,A^*+B\,B^*=I_m,\quad A^*\,A+C^*\,C=I_m,\quad
C\,C^*+D\,D^*=I_{n-m}\label{block1}
\end{eqnarray}
where in the following $I_k$ denotes the
$k\times k$ unit matrix.  Suppose we know the contraction $A$, then
the problem reduces to finding the  $B$, $C$ and $D$   blocks  such that $
A_n$
should be unitary. In other words the problem is: knowing a contraction $A$ of side $m$ how we
can border it for getting a unitary $n\times n$ matrix $A_n$? For solving this problem  we shall make use of the theory of  contraction operators.

An operator $T$ applying the Hilbert space $\cal{H}$ in the Hilbert space
$\cal{H}'$ is a contraction if for any $v\in
{\cal{H}}$,~~$||T\,v||_{{\cal{H'}}}\leq ||v||_{{\cal{H}}}$, i.e. $||T||\leq
1$, \cite{FN}. For any contraction we have $T^*\,T\leq I_{{\cal{H'}}}$ and
$T\,T^*\leq I_{{\cal{H}}}$ and the defect operators
\[
D_T=(I_{{\cal{H}}}-T^*\,T)^{1/2},\quad
D_{T^*}=(I_{{\cal{H'}}}-T\,T^*)^{1/2}
\]
 are Hermitean operators  in
${{\cal{H}}}$ and ${{\cal{H}'}}$ respectively. They have the property
\begin{eqnarray}
 T\,D_{T}=D_{T^*}\,T,\quad T^*\,D_{T^*}=D_{T}\,T^*\label{def}
\end{eqnarray}
Here we consider only finite-dimensional contractions,  i.e. $T$ will have in
general  $n_1$ rows and $n_2$ columns.

The unitarity relations (\ref{block1}) can be written as
\[
BB^*=D_{A^*}^2,\qquad C^*C=D_A^2
\]

According to Douglas lemma \cite{Do} there exist two contractions $U$ and $V$
such that 
\[
 B=D_{A^*}U,\quad {\rm and} \quad C=D_AV
\]
Since we are looking for a parameterisation of unitary matrices, $U^*$ and $V$
are isometries, i.e. they satisfy the relations
\[
UU^*=I_{n-m}, \qquad V^*V=I_{m}
\]
If $n$ is even and $m=n/2$, then $U$ and $V$ are unitary operators. Thus 
$B$ and $C$ blocks are given by the defect operators $D_{A^*}$, $D_A$ and
two arbitrary isometries whose dimensions are $m\times (n-m)$ and $(n-m)\times
m$ respectively.
  The last block of $A_n$ is given by the lemma

\newtheorem{Le}{Lemma }
\begin{Le}
The formula $$D=-VA^*U+D_{V^*}K D_U$$
establishes a one-to-one correspondence between all the bounded operators $D$
such that 
\[
A_n=\left( \begin{array}{cc}A &D_{A^*}U\\ VD_A&D\end{array}\right)
\]
is a contraction and all the bounded contractions $K$.
\end{Le}
 See Arsene and Gheondea \cite{AG} for a proof of the general result when $U$, $V$ and $K$ are
contractions, and further details. In our case $U$ and  $V$ being isometries
 $D$ is given by 
\begin{eqnarray}
D=-VA^*U + XMY\label{gheo}
\end{eqnarray}
where $X$ and $Y$ are those unitary matrices that diagonalise the Hermitean
defect operators $D_{V^*}$ and $D_U$ respectively, i.e.
\[
X^*D_{V^*}X=P,\qquad Y^*D_UY=P
\]
  $P$ is the projection
\[
P=\left( \begin{array}{cc}0 &0\\ 0&I_{n-2m}\end{array}\right)
\]
and the matrix $M$ entering (\ref{gheo}) has the form
\[
M=\left( \begin{array}{cc}0 &0\\ 0&A_{n-2m}\end{array}\right)
\]
where $A_{n-2m}$ denotes an arbitrary $(n-2m)\times(n-2m)$ unitary matrix.
See \cite{Di} for details. In the above formulae we supposed that
the eigenvectors of the  $D_U$ and $D_{V^*}$ operators entering the matrices $X$ and $Y$
are ordered in the increasing order of the eigenvalues.

Therefore the parameterisation of an $n\times n$ unitary matrix is equivalent
to the parameterisations of four matrix blocks with lower dimensions than those  of
the original matrix, and consequently our task is considerably simplified. On the
other hand the formulae (\ref{gheo}) and  subsequent show that this procedure is
recursive allowing the parameterisation of any finite dimensional unitary
matrix starting with the parameterisation of one- or two-dimensional unitary
matrices. Moreover the parameterisation of $A_n$ requires the parameterisation
of an $m\times m$ contraction, of two isometries $U$ and $V$ and of an
$(n-2m)\times(n-2m)$ unitary matrix. In our papers \cite{Di,Di2} we
considered only the case $m=1$ as the simplest one, however the case $m >1$ may be  useful in the study of complex Hadamard matrices. 

For what follows we treat again   the case  $m=1$, i.e. $A$ is the simplest contraction, a complex number whose modulus is less than one, because we found the form of  the matrices $X$ and $Y$
 for arbitrary $n$. Since $V$  is
a $(n-1)$-dimensional vector  the isometry property allows us to parametrise
it as $V=(cos\,\chi_1,sin\,\chi_1\, cos\,\chi_2,$
$\dots,sin\,\chi_1\dots
sin\,\chi_{n-2})^t$  where $t$ denotes transpose.  $V$ is the eigenvector of
$D_{V^*}$ corresponding to the zero eigenvalue. Indeed from the relations 
(\ref{def}) we
have $$D_{V^*}\,V=V\,D_{V}=0$$ showing that $V$ is the eigenvector of $D_{V^*}$
corresponding to the zero eigenvalue. Thus the problem is: how to complete an
orthogonal matrix $X$ knowing its first column (row) such that no suplementary
parameters enter. The other columns
of this matrix we are looking for will be given by  the other eigenvectors  of $D_{V^*}$. One easily verifies that $D_{V^*}$ is a projection operator such that the other eigenvalues  equal unity. Indeed the folowing holds

\setcounter{Le}{1}
\begin{Le}
The orthonormalised eigenvectors of the eigenvalue problem
\[
D_{V^*}\,v_k=\lambda_k\,v_k,\,\,\, k=1,\dots,n-1
\]
 are the columns of the orthogonal matrix $X\in SO(n-1)$ and are generated by
the vector $V$ as
\[
v_1=\left(\begin{array}{c}
cos\,\chi_1\\
sin\,\chi_1\, cos\,\chi_2\\
\cdot\\
\cdot\\
\cdot\\
sin\,\chi_1\dots
sin\,\chi_{n-2}
\end{array}\right)\]
and 
\[
v_{k+1}={d\over d\,\chi_k}\,v_1(\chi_1=\dots=\chi_{k-1}={\pi\over 2}),
\,\,k=1,\dots,n-2
\]
where in the above formula one calculates first the derivative and afterwards the
restriction  to $\pi/2$. \end{Le}
In a similar way one finds  $Y$; see \cite{Di3} for a proof.

In the case of $n \times n$ Hadamard matrices whose elements of the first row and of  the first column are positive numbers $a_{1j}=a_{j1}={1\over\sqrt{n}}$, $j=1,\dots,n$,  $X$ has the form
\[
\left(\begin{array}{cccccccc}
{1\over \sqrt{n-1}}&-\sqrt{n-2\over n-1}&0&0&\dots&\dots&0&0\\
{1\over \sqrt{n-1}}&{1\over \sqrt{(n-1)(n-2)}}&-\sqrt{n-3 \over n-2}&0&\dots&\dots&0&0\\
{1\over \sqrt{n-1}}&{1\over \sqrt{(n-1)(n-2)}}&{1\over\sqrt{(n-2)(n-3)}}&-\sqrt{n-4\over n-3}&\dots&\dots&0&0\\
\cdot&\cdot&\cdot&\cdot&\cdot&\cdot&\cdot&\cdot\\
\cdot&\cdot&\cdot&\cdot&\cdot&\cdot&\cdot&\cdot\\
\cdot&\cdot&\cdot&\cdot&\cdot&\cdot&\cdot&\cdot\\
{1\over\sqrt{n-1}}&{1\over \sqrt{(n-1)(n-2)}}&{1\over\sqrt{(n-2)(n-3)}}&{1\over\sqrt{(n-3)(n-4)}}&\dots&\dots& {1\over\sqrt{6}}&-{1\over\sqrt{2}}\\
{1\over\sqrt{n-1}}&{1\over \sqrt{(n-1)(n-2)}}&{1\over\sqrt{(n-2)(n-3)}}&{1\over\sqrt{(n-3)(n-4)}}&\dots&\dots& {1\over\sqrt{6}}&{1\over\sqrt{2}}\\
\end{array}\right)
\]
and $Y=X^t$, where $t$ denotes the transposed matrix.

In this way all the quantities entering formula (\ref{gheo}) are known and the
parameterisation of $A_n$ can be obtained recursively starting with the known
parameterisation of  $2\times 2$ unitary matrices.

  When the block $A$ is one-dimensional, i.e. a simple number equal to $1/\sqrt{n}$, the term
  $V\,A^*\,U$ entering Eq.(\ref{gheo})   has the form $\frac{1}{(n-1)\sqrt{n}}\,J$
  where $J$ is the $(n-1)\times(n-1)$ matrix whose each of entries is $+1$,
  which appears in many constructions of {\it real} Hadamard matrices; see
  Agaian \cite{Ag}.

\section{ Application}

In the following we  will use  Eq.(\ref{gheo}) to generalize to the case of complex
Hadamard matrices the trics used by Sylvester \cite{Sy} and Hadamard \cite{Ha}
for constructing complex Hadamard matrices. We take  $n$ an even number, $n=2\,
m$, and we suppose that we know a parameterisation of the $A$ block which is unitary  and whose order is $m$. In that case $B$ and $C$ blocks are also  unitary matrices of order $m$ and we consider them normalized as $A\,A^*=B\,B^*=C\,C^*= I_m$. From (\ref{gheo}) we have $D=-C\,A^*\,B$ and 
the following matrix
\[
{1\over\sqrt{2}}
\left(
\begin{array}{cc}
A&B\\
C&-C\,A^*\,B
\end{array}
 \right)
\]
will be  unitary by  construction. In general the above matrix will not be  Hadamard  even when $A,\,\,B$ and $C$ are, as the simplest example shows; this happens only when either $C=A$ or $B=A$. Since the second case is obtained by  transposing the  matrix of the first one,  as long as $B$ and $C$ are arbitrary, we will consider only the matrix
\begin{eqnarray}
 {1\over\sqrt{2}} \left( \begin{array}{cc}
A&B\\
A&-B
\end{array} \right)\label{arr}
\end{eqnarray}
which is the elementary two-dimensional array that will be used in the construction of more complicated arrays of Hadamard matrices.
In the following we suppose that $A$ and $B$ are complex Hadamard matrices of size $m$ each one depending on $p\ge 0$, respectively, $q\ge 0$ free phases, i.e. (\ref{arr}) is a complex Hadamard matrix of size $2\,m$. Now we make use of Hadamard's trick to get a Hadamard matrix depending on $p+q+m-1$ arbitrary phases. Indeed we can multiply $B$ at left by the diagonal matrix $d=(1,e^{i\,\varphi_1},\dots,e^{i\,\varphi_{m-1}})$ without modifying the Hadamard property. In this way Hadamard obtained a continuum of solutions for the case $n=4$. We denote $B_1=d\cdot B$ and then the matrix
\begin{eqnarray}
\frac{1}{\sqrt{2}}
\left(
\begin{array}{cc}
A&B_1\\
A&-B_1
\end{array}
 \right)\label{arr1}
\end{eqnarray}
will be unitary and Hadamard depending on $p+q+m-1$ parameters. From (\ref{arr1}) we obtain in general
two non-equivalent $2\,m\times 2\, m$ Hadamard matrices when $B\neq B^*$.  In this case  Eq.(\ref{arr1}) is a realization and the second one is given by $B_1 \rightarrow B_2 =d\cdot B^*$. 
The above procedure can be iterated by taking 
the matrix (\ref{arr}) as a new $A$ block obtaining a Hadamard matrix of the form
\begin{eqnarray}
\frac{1}{2}
\left(
\begin{array}{crrr}
A&B&C&D\\
A&-B&C&-D\\
A&B&-C&-D\\
A&-B&-C&D
\end{array}
 \right)\label{arr2}
\end{eqnarray}
which is a $4\,m$-dimensional array similar to Williamson array \cite{Wi}, and
so on. 
 In contradistinction to the Williamson array the $A,\, B,\, C,\, D$
blocks satisfy no supplementary conditions, excepting their unitarity.
Thus the following holds
\newtheorem{pro}{Proposition}
\begin{pro}
If the $m\times m$ complex Hadamard matrices $A, B, C, D$ depend on $p, q, r,s$
arbitrary phases then there exists a complex Hadamard matrix of the form (\ref{arr2}) which depends on $p+q+r+s+3(m-1)$ arbitrary phases.
\end{pro}

 We
notice that the elementary array (\ref{arr}) is different from the Goethals-Seidel one \cite{GS}
that appears  in the  construction of {\it real} Hadamard matrices and which  has the form
\[
\frac{1}{\sqrt{2}} \left( \begin{array}{cc}
A&B\\
B&-A
\end{array} \right)
\]
The above array is not unitary even when $A$ and $B$ are, the suplementary
condition for unitarity being the relation  $A\,B^*=B A^* $. We consider that the form (\ref{arr}) could also be useful  for the study of orthogonal designs and {\it real} Hadamard matrices it being in some sense complementary to the above form.

As an application of the formula (\ref{arr2}) we consider the following case: $a_{11}=a_{12}=a_{21}=-a_{22}=b_{11}=b_{12}=c_{11}=c_{12}=d_{11}=d_{12}=1/\sqrt{2}$ and
$b_{21}=-b_{22}=e^{is}/\sqrt{2}$, $c_{21}=-c_{22}=e^{it}/\sqrt{2}$, $d_{21}=-d_{22}=e^{iu}/\sqrt{2}$ where the notation is self-explanatory, and we obtain an eight-dimensional Hadamard matrix depending on three arbitrary phases $s,\,t,\,u$.

 When $A=B$, Eq.(\ref{arr}) can be written as
\begin{eqnarray}
 \frac{1}{\sqrt{2}} \left( \begin{array}{cc}
A&A\\
A&-A
\end{array} \right)={1\over\sqrt{2}} \left( \begin{array}{cc}
1&1\\
1&\epsilon
\end{array} \right)\otimes A \label{syl}
\end{eqnarray}
where $\epsilon =-1$, i.e. the first factor is the Sylvester Vandermonde matrix of the second roots of unity,   and $\otimes$ is the ordinary Kronecker product, $A\otimes B=[a_{ij}B]$; of course the first factor can be any complex  Hadamard matrix of order $m$. Now we want to define a new product the aim being a more general construction of Hadamard matrices. Let  $M$ and $N$ be two matrices of the same order $m$ whose elements are matrices $M_{ij}$ of order $n$ and respectively $N_{kl}$ of order $p$. The new product denoted by $\tilde\otimes$ is given as
\[
Q=M\tilde{\otimes}N
\]
which is a matrix of order $mnp$, where
\[
Q_{ij}=\sum_{k=1}^{k=m}\,M_{ik}\otimes N_{kj}
\]
We will use here the above formula only in the  case: 
 $M=m_{ij}$ where $m_{ij}$ are complex scalars, not matrices and $N$ is an arbitrary diagonal  matrix $N=(N_{11},\cdots, N_{mm})$ where $N_{ii}$ ar matrices  of order $p$ obtaining 
\begin{eqnarray}
 Q= \left( \begin{array}{cccc}
m_{11}N_{11}&\cdot&\cdot&m_{1m}N_{mm}\\
\cdot&\cdot&\cdot&\cdot\\
\cdot&\cdot&\cdot&\cdot\\
m_{1m}N_{11}&\cdot&\cdot&m_{mm}N_{mm}
\end{array} \right)\label{arr3}
\end{eqnarray}
Thus the following is true.
\setcounter{pro}{1}
\begin{pro}
If the matrices $M$ and $N_{ii},\, i=1,\dots,m$, are Hadamard so will be the
matrix  $Q$ given by Eq.(\ref{arr3}).
\end{pro}
 The order of $Q$ is $mp$ and the formula (\ref{arr3}) is new even for real
Hadamard matrices.
 This form is the most general array we have
obtained and in some sense (\ref{arr3})  is the natural generalization of Williamson arrays to the case of complex Hadamard matrices.

    If in the above relation  we take $m_{11}=m_{12}=m_{21}=-m_{22}={1/ \sqrt{2}}$ and   $N_{11}=A$ and $N_{22}=B$, then Eq.(\ref{arr3}) reduces to Eq.(\ref{arr}).

\begin{newtheorem}{ex}{Example}
\begin{ex}

 If now $m_{ij}$ are the same as above  and
$$N_{11}={1\over 2}\left( \begin{array}{rrrr}
1&1&1&1\\
1&1&-1&-1\\
1&-1&-e^{is}&e^{is}\\
1&-1&e^{is}&-e^{is}
\end{array}\right)$$
is the complex four-dimensional Hadamard matrix and

$$N_{22}={1\over 2}\left( \begin{array}{rrrr}
1&0&0&0\\
0&e^{it}&0&0\\
0&0&e^{iu}&0\\
0&0&0&e^{iv}
\end{array}\right)
\left( \begin{array}{rrrr}
1&1&1&1\\
1&1&-1&-1\\
1&-1&-e^{iy}&e^{iy}\\
1&-1&e^{iy}&-e^{iy}
\end{array}\right)$$
we obtain an eight-dimensional matrix depending now on five arbitrary phases
$s,t,u,v,y$ instead of three as in the preceding example obtained by using the
Williamson-type array (12).
\end{ex}
\end{newtheorem}
 Thus the following holds.
\setcounter{pro}{2}
\begin{pro}
 If $M,N_i,\,i=1,\dots,m$ are $m\times m$ and respectively, $ n\times n$-dimensional complex Hadamard matrices depending on $m$, respectively, $n_i$, arbitrary phases, then there is an array of the form (14) that depends on
\[
m+n_1+(m-1)\sum_{i=2}^m m_i
\]
free phases.
\end{pro}

The above example shows the necessity for getting upper and lower bounds on the number of arbitrary  phases entering a Hadamard matrix of size $N$. Taking into account the standard decomposition of any integer under the form $N=p_1^{q_1}\dots p_m^{q_m}$, where $p_1 <\dots <p_m$ are primes and  $q_1 \dots q_m$ their respective powers, we may use the above {\em Proposition 3} for obtaining lower bounds on the number of free phases, that  we shall denote it by $\varphi(N)$. Since until now does not exist an example of a Hadamard matrix of size $N$ with $N$ prime which depends on free phases, in the following we will consider the normalization  $\varphi(N)=0$, for $N$ prime. Thus  the following holds.

\setcounter{so}{1}
\begin{so}
Let $N=p_1^{q_1}$ be the power of a prime $p_1$, with $q\ge 2$. Then a lower bound for $\varphi(p_1^{q_1})$, the number of free phases entering the parameterization of the $N\times N$ complex Hadamard matrix, is given by
$$ \varphi(p_1^{q_1})= 1+[(p_1-1)(q_1-1) -1]p_1^{q_1 -1}$$

If $N=p_1^{q_1}\dots p_m^{q_m}=p_1^{q_1} N_1 $ then $\varphi(p_1^{q_1}N_1)$ is given by
$$\varphi(p_1^{q_1}N_1)=1+[(p_1 -1)q_1N_1-p_1]p_1^{q_1 -1} + \varphi(N_1)p_1^{q_1}$$
\end{so}

{\em Proof.} Making use of  {\em Proposition 3} we find the recurrence relation
$$
 \varphi(p_1^{q_1})= p_1 \varphi(p_1^{q_1 -1})+(p_1-1)(p_1^{q_1 -1}-1)
$$
 with the initial condition $\varphi(p_1)=0$ and the solution follows.

In the second case the recurrence relation writes
\[
\varphi(p_1^{q_1}N_1)=p_1\varphi(p_1^{q_1 -1} N_1)+(p_1 -1)(p_1^{q_1 -1} N_1 -1)
\]
and the initial condition can be taken as 
$$\varphi(p_1 N_1)= p_1\varphi(N_1)+
(p_1 -1)(N_1 -1)$$
 and the solution follows. The above recurrence relation allows us to obtain lower bounds for any integer $N$ under the form
$$
\varphi(p_1^{q_1}\dots p_m^{q_m})=1+ [(p_1 -1)q_1p_2^{q_2}\dots p_m^{q_m} -p_1]p_1^{q_1 -1}+
$$
$$
p_1^{q_1}\{1+[(p_2 -1)q_2 p_3^{q_3}\dots p_m^{q_m} -p_2]p_2^{q_2 -1}+
$$ 
$$
p_2^{q_2}\{1+[(p_3 -1)q_3 p_4^{q_4}\dots p_m^{q_m} -p_3]p_3^{q_3 -1}+
$$
$$
p_3^{q_3}\{1+\dots +p_{m-1}^{q_{m-1}}\{1+[(p_m -1)q_m -p_m]p_m^{q_m -1}\}+
$$
$$
p_{m-1}^{q_{m-1}}\{1+[(p_m -1)(q_m -1) -1]p_m^{q_m -1}\}\dots\}
$$
We give now a few examples.
\setcounter{ex}{1}
\begin{ex}
 If $N=p_1^{q_1}p_2^{q_2}$ then
  the lower bound for
$\varphi(p_1^{q_1}p_2^{q_2})$,  the  number of  free phases entering the parameterization of the $N\times N$  complex Hadamard matrix, is given by
\begin{eqnarray}
\varphi(p_1^{q_1}p_2^{q_2})= 1+ (p_1 -1)q_1p_1^{q_1 -1}p_2^{q_2}+ [(p_2 -1)(q_2 -1)- 1]p_1^{q_1}p_2^{q_2 -1}\label{bound1}
\end{eqnarray}
\end{ex}
Numerical examples:
$\varphi(2^3)=5,\, \varphi(2^4)=17,\,\varphi(6)=
2,\, \varphi(3^2)=4,\varphi(2^2 3^2)=49,$ etc.

\section{An other parameterisation of unitary matrices}

In the following we will shortly present  another parameterisation of unitary matrices \cite{Di3}
under the form of a product of $n$ diagonal matrices containing phases
interlaced with $n-1$ orthogonal matrices each one generated by a real vector
$v\in {\bf R}^n$. This new form will be more appropriate for design and
implementation of the  software packages necessary for solving  the equations (\ref{mod}) for arbitrary $n$.  

We have seen in Section 2 that we can write any unitary matrix as a product of two diagonal matrices of the for $d_n=(e^{i\varphi_1},\dots,e^{i\varphi_n})$ with $\varphi_j \in [0,2\,\pi)$, $j=1,\dots,n$ arbitrary phases and a unitary matrix with positive elements in the first row and the first column. 
We make also the notation $d_k^{n-k}=(1_{n-k}, e^{i\psi_1},\dots,e^{i\psi_k})$,\, $k<n$, where $1_{n-k}$ means  that the first $(n-k)$ diagonal entries  equal  unity,
i.e. it can be obtained  from $d_n$ by making the first $n-k$ phases equal zero. These diagonal phase matrices are the first building blocks in our construction.
 Other  building blocks that will appear in factorization of unitary matrices  $A_n$ 
are the two-dimensional rotations which operate in the $i,i+1$-plane of the form 
\begin{eqnarray}
J_{i,i+1}=\left(
\begin{array}{ccc}

I_{i-1}&0&0\\
0&
\begin{array}{cc}
cos\,{\theta_i}&-sin\,{\theta_i}\\
sin\,{\theta_i}& cos\,{\theta_i}
\end{array}
&0\\
0&0&I_{n-i-1}
\end{array}\right),\quad i=1,\dots,n-1\label{rot} 
\end{eqnarray}

The factorization idea comes from the well known fact that $U(n)$ acts transitivly on the $n$-dimensional complex sphere ${\bf S}_{2n-1}\,\in{\bf C}^{n}$, and explicitely from the coset relation

\[
{\bf S}_{2n-1}={\it coset~space }\,\,{ U}(n)/{ U}(n-1)
\]

A
direct consequence of the last relation is that  we expect that any element of
${ U}(n)$ should be uniquely specified by a pair of a vector $v\in {\bf S}_{2n-1}$ and
of an arbitrary element of ${ U}(n-1)$. Thus we are looking for a factorization
of an arbitrary element $A_n\in { U}(n)$ in the form
\[
A_n=B_n\cdot\left(\begin{array}{cc}
1&0\\0&A_{n-1}\end{array}\right)
\]
 where $B_n\in{ U}(n)$ is a unitary
matrix whose first column is uniquely defined by a vector $v\in {\bf S}_{2n-1}$, but
otherwise   arbitrary,  and $A_{n-1}$ is an arbitrary element of ${ U}(n-1)$.
Iterating the previous equation we arrive at the conclusion that an element of
${ U}(n)$ can be written as a product of $n$ unitary matrices 
\[                             
A_n=B_{n}\cdot B_{n-1}^1\dots B_1^{n-1}
\]
where
\[
B_{n-k}^k=\left(\begin{array}{cc}
I_k&0\\
0&B_{n-k} \end{array}\right)
\]
$B_k,\,\, k=1,\dots,n-1$, are $k\times k$ unitary matrices whose
first columns are generated by vectors $b_k\in {\bf S}_{2k-1}$; for
example $B_1^{n-1}$ is the diagonal matrix
$(1,\dots,1,e^{i\varphi_{n(n+1)}})$.
 
The still arbitrary columns of $B_k$ will be chosen in such a way that we
should obtain a simple form for the matrices $B^{n-k}_k$, and we 
require that  $B_k$ should be completely specified by the parameters entering
the vector $b_k$ and nothing else.

If we take into account the equivalence considerations of the  Section 2 then
$B_n\,(B_{n-k})$ can be written as
\[
B_n=d_n\,\tilde{B_n}
\]
where the first column of $\tilde{B_n}$ has non-negative entries.

 Denoting
this column by $v_1$ we will use the parameterization 
\[
v_1=(cos\,{\theta_1},cos\,{\theta_2}\,sin\,{\theta_1
},\dots,\sin\,{\theta_1}\dots sin\,{\theta_{n-1}})^t
\]
where $\theta_i\in[0,\pi/2],\, i=1,\dots,n-1$. Thus
$B_n$ will be parameterized by $n$ phases and $n-1$ angles. According to the
above factorization $\tilde{B}_n$ is nothing else than the orthogonal matrix
generated by the vector $v_1$ and its form is given by {\em Lemma 2 }with $n\rightarrow n+1$. Thus without loss of generality
$B_n=d_n\,{\cal{O}}_n$ with ${\cal{O}}_n\in SO(n)$.
In this way the factorization of $A_n$ will be
\begin{eqnarray}
A_n=d_n\,{\cal{O}}_n\,d_{n-1}^1\,{\cal{O}}_{n-1}^1\dots
d_2^{n-2}{\cal{O}}_{2}^{n-1}d_1^{n-1}I_n \label{fac} 
\end{eqnarray}
where ${\cal{O}}_{n-k}^k$ has the same structure as $B_{n-k}^k$, i.e
\[
 {\cal{O}}_{n-k}^k=\left(\begin{array}{cc}
I_k&0\\
0&{\cal{O}}_{n-k}
\end{array}\right)
\]
and $d^k_{n-k}=(1,\dots,1,e^{i\phi_1},\dots,,e^{i\phi_{n-k}})$

The orthogonal matrices ${\cal{O}}_n$ can be factored in terms of $J_{i,i+1}$ as follows.

\setcounter{Le}{2 }
\begin{Le}

  The orthogonal matrices ${\cal O}_n$ ( ${\cal O}_{n-k}^k$)
at their turn can be  factored into a product of $n-1$ (n-k-1)
matrices of the form $J_{i,i+1}$; e.g. we have 
\begin{eqnarray}
{\cal O}_n=J_{n-1,n}\,J_{n-2,n-1}\dots
J_{1,2}\nonumber
\end{eqnarray}
where $J_{i,i+1}$ are $n\times n$ rotations introduced by Eq.(\ref{rot}). 
\end{Le}

 In this way the parameterisation of unitary matrices reduces to a product of simpler matrices: diagonal phase matrices and two-dimensional rotation matrices. For more details see our paper \cite{Di3}.  Now we propose a disentanglement of the angles and phases  entering each ``generation'' and denote   the angles by latin letters, e.g.  those that parameterize ${\cal O}_n$ will be denoted by
$a_1,\dots,a_{n-1}$,  the angles that parameterize ${\cal
O}_{n-1}^1$, by $b_1,\dots,b_{n-2}$, etc., the
last angle entering ${\cal
O}^{n-1}_2$ by $z_{1}$. The phases will be denoted by Greek letters; e.g. the phases entering $d_1$ will be denoted by  $\alpha_1,\dots,\alpha_{n}$, those entering $d_{n-1}^1$ by $\beta_1,\dots,\beta_{n-1}$, etc. The above factorization will be used in the next section for obtaining the equations for the moduli of the matrix elements.

\section{Explicit equations of the  moduli}

Our  choise for the orthogonal vectors in {\it Lemma 2} was such that the resulting matrix should have as many zero entries  as possible. Thus ${\cal{O}}_n$ has $(n-1)(n-2)/2$ zeros in the right upper corner and the entries of the Hadarmard matrix will get more and more complicated when going from left to right and from top to bottom. We will start using the form (\ref{fac}) of the unitary matrix and then 
$d_n\equiv I_n$. Since the first column has the form $a_{i1}=1/\sqrt{n},\,\, i=1,\dots,n$ and $d_{n-1}^1=(1,e^{i\alpha},e^{i\alpha_1},\dots,e^{i\alpha_{n-2}})$ the product ${\cal{O}}_n\,d_{n-1}^1$ is
$$
\left(\begin{array}{ccccccc}
{1\over \sqrt{n}}&-\sqrt{{n-1\over n}}\,e^{i\alpha}&0&0&\dots&0&0\\

{1\over \sqrt{n}}&{e^{i\alpha}\over \sqrt{n(n-1)}}&-\sqrt{{n-2 \over n-1}}\,e^{i\alpha_1}&0&\dots&0&0\\

{1\over \sqrt{n}}&{e^{i\alpha}\over \sqrt{n(n-1)}} &{e^{i\alpha_1}\over\sqrt{(n-1)(n-2)}}&-\sqrt{{n-3\over n-2}} e^{{i\alpha_2}}&\dots&0&0\\
\cdot&\cdot&\cdot&\cdot&\cdot&\cdot&\cdot\\
\cdot&\cdot&\cdot&\cdot&\cdot&\cdot&\cdot\\
\cdot&\cdot&\cdot&\cdot&\cdot&\cdot&\cdot\\

{1\over \sqrt{n}}&{ e^{i\alpha}\over \sqrt{n(n-1)}}& e^{i\alpha_1}\over\sqrt{(n-1)(n-2)}&e^{i\alpha_2}\over\sqrt{(n-2)(n-3)} &\dots& e^{i\alpha_{n-3}} \over\sqrt{6}&-e^{i\alpha_{n-2}} \over\sqrt{2}\\

{1\over \sqrt{n}}&{e^{i\alpha}\over \sqrt{n(n-1)}}&{e^{i\alpha_1} \over\sqrt{(n-1)(n-2)}}&{e^{i\alpha_2} \over\sqrt{(n-2)(n-3)}}&\dots&{ e^{i\alpha_{n-3}} \over\sqrt{6}}&{e^{i\alpha_{n-2}} \over\sqrt{2}}\\
\end{array}\right) \eqno(18)$$
where $\alpha, \alpha_i,\, i=1,\dots,n-2$ are $n-1$ arbitrary phases.

The next building block ${\cal{O}}_{n-1}^1\,d_{n-2}^2$   will have the form
$$
\left(\begin{array}{ccccc}
1&0&0&\cdot&0\\
0&\cos\,a&-\sin\,a\,e^{i\beta}&\cdot&0\\
0&\sin\,a\, \cos\,a_1&\cos\,a \,\cos\,a_1\,e^{i\beta}&\cdot&0\\
\cdot&\cdot&\cdot&\cdot&\cdot\\
\cdot&\cdot&\cdot&\cdot&\cdot\\
0&\sin\,a...\sin\,a_{n-3}&\cos\,a\,\sin\,a_1\cdots \sin\,a_{n-3}\,e^{i\beta}&\cdot&
\cos\,a_{n-3}\,e^{i\beta_{n-3}}\\
\end{array}
\right)\eqno(19)
$$
in terms of $n-2$ phases $\beta,\beta_1,\dots,\beta_{n-3}$ and $n-2$ angles $a,a_1,\dots, a_{n-3}$, and so on.

It is easy to see that the first two columns of the product of matrices (18) and (19) does not change when multiplied by ${\cal O}_{n-2}^2\,d_{n-3}^3$; however the first row does. If the angles entering  ${\cal O}_{n-2}^2$ are denoted by $b, b_1,\dots,b_{n-4}$ and the phases are $\gamma,\gamma_1,\dots, \gamma_{n-4}$, etc., then the entries of the first row are
\[
 a_{12}=-\sqrt{ n-1\over n}\cos\,a\,e^{i\alpha},\quad a_{13}=\sqrt{ n-1\over n}\sin\,a \,\cos\,b\,e^{i(\alpha+\beta)}, \dots,
\]
\[
 a_{1 n-1}=(-1)^{n-1}\sqrt{ n-1\over n}\sin\,a \,\sin\,b\, \dots \cos\,z\, e^{i(\alpha+\beta+ \dots \omega)}
\]
where $z\,\, {\rm and}\,\, \omega$ are the last angle and phase respectively.
Since we use the standard form of Hadamard matrices, i.e. the entries of the first row and of the first column are positive and equal $1/\sqrt n$, the above equations imply
$$\alpha=\beta=\dots =\omega=\pi; \,\, \cos\, a={1\over\sqrt{n-1}},\, \cos\, b={1\over\sqrt{n-2}},\dots, \cos\, z={1\over\sqrt{2}}$$
We substitute the above values in Eq.(\ref{fac}) and find a complex $n\times n$ matrix depending on $(n-1)(n-2)/2$ phases $\alpha_1,\dots,\alpha_{n-2},\beta_1,\dots,\psi_1$ and $(n-2)(n-3)/2$ angles $a_1,\dots,a_{n-3},b_1,\dots,y_1$, i.e. $(n-2)^2$ parameters which have to be found by solving the corresponding equations given by the moduli. The first simplest  entries of the unitary matrix  have the form

\[
  a_{22}=-{1\over{(n-1)}\sqrt{n}}-{n-2\over n-1}\cos\,a_1\,e^{i\alpha_1},\dots
\]
$$
a_{k2}=-{1\over{(n-1)}\sqrt{n}}\, +\,\sqrt{{n-2\over n-1}}\,\left({\cos\,a_1\,e^{i\alpha_1}\over\sqrt{(n-1)(n-2)}}+\dots+ { \sin\,a_1\dots \cos\,a_{k-2}\,e^{i\alpha_{k-2}}\over\sqrt{(n-k+2)(n-k+1)}}\right.
$$
$$
-\left.{\sqrt{{n-k\over n-k+1}}}\sin\,a_1\dots \sin\,a_{k-2}\cos\,a_{k-1}\,e^{i\alpha_{k-1}}\right),\,\, k=3,\dots,n-1\eqno(20)
$$
$$
a_{2k}=-{1\over{(n-1)}\sqrt{n}}\,+\,\sqrt{{n-2\over n-1}}\,\left({\cos\,a_1\,e^{i\alpha_1}\over\sqrt{(n-1)(n-2)}}-{\sin\,a_1\cos\,b_1\,e^{i(\alpha_1+\beta_1)}\over\sqrt{(n-2)(n-3)}}\,+\,\dots\right.
$$
$$
\left. +(-1)^{k-1}\,\,  \sqrt{n-k\over {n-k+1}}\,\sin\,a_1\sin\,b_1\dots\cos\,l(k)_1\,e^{i(\alpha_1 + \beta_1 +\dots +\lambda(k)_1)}\right),\,\,{\rm etc.}$$
where $l(k)\,{\rm and}\,\lambda(k)$ denote the letters for angle and respectively phase corresponding to index $k$ and the signs in the last bracket alternate.

The matrix elements get more complicated when  going from the upper left corner to right bottom corner. The entries $a_{22}, a_{32}$  and $ a_{23}$ lead, for example,  to the following moduli equations
\[
(n-2)\,\cos^2 a_1\,+\,{2\over\sqrt{n}}\,\cos\,a_1\, \cos\,\alpha_1\,-\,1=0
\]
\[
\sin a_1\left((n-3)\sin a_1\,\cos^2 a_2 +\right.
\]
$$~~~~\left. 2 \sqrt{n-3\over n-1}\cos a_2\left({\cos \alpha_2\over\sqrt{n}} - \cos a_1\,\cos(\alpha_1-\alpha_2)\right) - \sin\,a_1 \right)=0\eqno(21)$$
\[
\sin a_1\left((n-3)\sin a_1\,\cos^2 b_1 +\right.
\]
\[
\left. 2 \sqrt{n-3\over n-1}\cos b_1\left(-{\cos(\alpha_1 + \beta_1)\over\sqrt{n}} + \cos a_1\,\cos \beta_1 \right) - \sin\,a_1 \right)=0
\]
and so on. The form of the last two equations was obtained after the elimination of  the term containing $\cos a_1\, \cos \alpha_1$ by using the first equation (21), i.e. we work in the ideal generated by the moduli equations. It is easily seen that the other equations contain as factors $\sin a_2,\dots, \sin a_{n-2}, \sin b_{1},\dots,{\rm etc.}$. Thus a particular solution can be obtained when
\[
\sin a_1 =0
\]
 which implies $ a_1=0,\pi$, and from the first equation  (21) we get
\[
\cos \alpha_1=\pm {(n-3)\sqrt{n}\over 2}
\]
It is easily seen that the above equation has solution only for $n=2,3,4$; for $n\ge 5$ the factor $\sin a_1$ will be omitted from Eqs.(21) because then $a_1\ne 0,\pi$. When $n=2$ we obtain $\alpha_1=\pi/4$, so $a_{22}=-1/\sqrt{2}$.  If $n=3$, then $\alpha_1=3 \pi/2$ and from the first Eq.(20)  one gets
\[
a_{22}=-{1\over 2\sqrt{3}}+{i\over 2}={1\over\sqrt{3}} e^{{2  \pi i\over 3}},\,\,{\rm etc.}
\]
The case $n=4$ leads to $\alpha_1=\pi$ which gives 
\[
a_{22}=-a_{23}=-a_{32}={1\over 2} \quad{\rm and} \quad a_{33}=-a_{34}=-{e^{i(\alpha_2 + \beta_1)}\over 2}
\]
After the substitution $\alpha_2 + \beta_1=t$ one finds the standard complex  form of the $4\times 4$  matrix  found  by Hadamard.
To view what is the origin of the phase $\alpha_2 + \beta_1$ we have to look at the moduli equations.
They  have the form
\[
2 \cos^{2}a_1 +\cos a_1\cos{\alpha_1} -1=0
\]
\[
\sin a_1(\cos{\alpha_2}- 2\cos{a_1}\cos(\alpha_1-\alpha_2))=0
\]
\[
\sin a_1(2 \cos{a_1}\cos{\beta_1}-\cos(\alpha_1+\beta_1))=0
\]
$$\cos 2a_1\, \cos(\alpha_1-\alpha_2)\cos\,\beta_1+\cos a_1\cos(\alpha_2+\beta_1)+ \sin(\alpha_1-\alpha_2)\sin \beta_1=0$$
and we see that   the above system splits into two cases. In the first case, when $\sin a_1=0$,  the rank of the system  is two which explains the above dependence of $a_{33}$ on two phases  and in the second case when  $\sin a_1\ne 0$  the rank is three and the dependence is only on one arbitrary phase. However in this case there is no final difference between the two cases. The solution of the above system is obtained directly but for $n\ge 5$ the problem is difficult and needs more powerful techniques. Particular solutions can be obtained rather easily e.g for $n=6$ there is a matrix that has the property $a_{ij}=a_{ji}$. 
\[
{1\over\sqrt{6}}\left(\begin{array}{rrrrrr}
1&1&1&1&1&1\\
1&-1&-1&1&i&-i\\
1&-1&-i&-1&1&i\\
1&1&-1&-i&-1&i\\
1&i&1&-1&-1&-i\\
1&-i&i&i&-i&-1
\end{array}\right)
\]
 There exists even a Hermitian matrix $S=S^*$

\[
{1\over\sqrt{6}}\left(\begin{array}{rrrrrr}
1&1&1&1&1&1\\
1&-1&i&i&-i&-i\\
1&-i&-1&1&-1&i\\
1&-i&1&-1&i&-1\\
1&i&-1&-i&1&-1\\
1&i&-i&-1&-1&1
\end{array}\right)
\]
 and so on. As we said before getting the most general form of a solution is not a simple task; for $n=6$ we have $16$ complicated trigonometric  equations and we remind that the simpler $ (\ref{sys})$ system was solved only for $n\le 8$ equations. Thus new approaches are necessary and in the next Section we suggest such an approach:  using methods from  algebraic geometry. 

\section{Connection with algebraic geometry}

  The Eqs.(21) can be transformed into polynomial equations by the known procedure 
\[
\sin\,a \rightarrow {2\,x\over 1+x^2}\,,\quad \cos\, a \rightarrow {1-x^2\over 1+x^2}
\]
such that we get  from (21)
$$p_1=\left[(n - 3+{2\over\sqrt{n}})x_1^4 -2(n-1)x_1^2+(n - 3-{2\over\sqrt{n}})\right]y_1^2+(n - 3-{2\over\sqrt{n}})x_1^4 -
$$
\[ 2(n-1)x_1^2+(n - 3+{2\over\sqrt{n}})  
\]                                                                             $$p_2=\left\{\left[-(1-{1\over\sqrt{n}})x_1^2+C_1\,x_1 +(1+{1\over\sqrt{n}})\right]x_2^4-C_2\,x_1\,x_2^2  +(1-{1\over\sqrt{n}})x_1^2+C_1x_1-\right.$$
$$\left.(1+{1\over\sqrt{n}})\right\}y_1^2y_2^2+ \left\{\left[(1-{1\over\sqrt{n}})x_1^2+C_1x_1-(1+{1\over\sqrt{n}})\right]x_2^4-C_2\,x_1\,x_2^2-(1-{1\over\sqrt{n}})x_1^2+\right.$$
$$\left.C_1\,x_1+ (1+{1\over\sqrt{n}})\right\}y_1^2+
 \left\{\left[(1+{1\over\sqrt{n}})x_1^2+C_1x_1-(1-{1\over\sqrt{n}})\right]x_2^4-C_2\,x_1\,x_2^2-\right.$$
$$\left.(1+{1\over\sqrt{n}})x_1^2+C_1\,x_1
 +(1-{1\over\sqrt{n}})\right\}y_2^2-4(1-x_1^2)(1-x_2^4)y_1y_2+\left[-(1+{1\over\sqrt{n}})x_1^2 +\right.
$$
$$
\left.C_1\,x_1
 +(1-{1\over\sqrt{n}})\right]x_2^4-C_2\,x_1\,x_2^2+(1+{1\over\sqrt{n}})x_1^2+C_1x_1-(1-{1\over\sqrt{n}}) \eqno(27)  $$                                                                                                                        $$p_3=\left\{\left[-(1-{1\over\sqrt{n}})x_1^2+C_1\,x_1 +(1+{1\over\sqrt{n}})\right]x_3^4-C_2\,x_1\,x_3^2  +(1-{1\over\sqrt{n}})x_1^2+C_1x_1-\right.$$
$$\left.(1+{1\over\sqrt{n}})\right\}y_1^2y_3^2+ \left\{\left[(1-{1\over\sqrt{n}})x_1^2+C_1x_1-(1+{1\over\sqrt{n}})\right]x_3^4-C_2\,x_1\,x_3^2-(1-{1\over\sqrt{n}})x_1^2+\right.$$
$$\left.C_1\,x_1+ (1+{1\over\sqrt{n}})\right\}y_1^2+
 \left\{\left[-(1-{1\over\sqrt{n}})x_1^2+C_1x_1+(1-{1\over\sqrt{n}})\right]x_3^4-C_2\,x_1\,x_3^2+\right.$$
$$\left.(1+{1\over\sqrt{n}})x_1^2+C_1\,x_1
 -(1-{1\over\sqrt{n}})\right\}y_3^2-4(1+x_1^2)(1-x_3^4)y_1y_2+\left[(1+{1\over\sqrt{n}})x_1^2 +\right.
$$
$$\left.C_1\,x_1
 -(1-{1\over\sqrt{n}})\right]x_3^4-C_2\,x_1\,x_3^2-(1+{1\over\sqrt{n}})x_1^2+C_1x_1+(1-{1\over\sqrt{n}})\eqno(26)   $$
where 
\[C_1={(n-1)(n-4)\over\sqrt{(n-1)(n-3)}}, \qquad C_2={2(n-1)(n-2)\over\sqrt{(n-1)(n-3)}}
\]                                                                  
  and the angles by the above transformation go  to $x_1,x_2,x_3,\dots$ and the phases to   $y_1,y_2,y_3,\dots$

From the matrices such as  (18) one sees that  the full set of the $(n-2)^2$ equations  contains  square roots of almost all prime numbers\,  $\le n$ so that not all the coefficients are  rational and we have to look for solutions in a field ${\mathbf{Q}}(\sqrt{d})$ for some $d\in{\mathbf{N}}$.                  

The polynomial equation $p_1=0$ defines an algebraic curve; however the most studied are the elliptic and hyperelliptic curves, i.e.  those defined by an equation of the form
$ y^2=f_p(x)$ where $f_p(x)$ is a polynomial of degree $p$. 

From $p_1=0$ we get
\[
y_1^2=-{(n - 3-{2\over\sqrt{n}})x_1^4
 -2(n-1)x_1^2+(n - 3+{2\over\sqrt{n}})\over (n - 3+{2\over\sqrt{n}})x_1^4 -2(n-1)x_1^2+(n - 3-{2\over\sqrt{n}})}=-{P_1(x_1)\over P_2(x_1)}
\]
 which defines a meromorphic function. 
Its zeros and poles are
\[
 \pm \sqrt{{\sqrt{n}-1\over \sqrt{n}+1}},\qquad \pm \sqrt{{n+\sqrt{n}-2\over
n-\sqrt{n}-2}}
\] and 
\[
\pm \sqrt{{\sqrt{n}+1\over \sqrt{n}-1}},\qquad \pm \sqrt{{n-\sqrt{n}-2\over
n+\sqrt{n}-2}}
\]
 respectively that are simple, and the poles and the zeros are interlaced. Thus apparently the above equation is not hyperelliptic, however by the birational transformation
\[
y_1={Y_1\over P_2(x_1)}
\]
we get the equation
\[
Y_1^2=-P_1(x_1)\,P_2(x_1)
\]
which shows that the above curve has genus $g=3$. For $n\ge 5$ the curve has no branch going to infinity since the highest power coefficient is negative and consequently the curve is made of three ovals.
 
The polynomials $p_1=p_2=0$ define a surface, $p_1=p_2=p_3=0$ define a 3-fold, and so on. We consider that the study of these multi-fold varieties will be very interesting from the algebraic geometry point of view and their parameterizations  could reveal unknown properties that may lead to a better understanding of the rational varieties. As we saw in Section 5 one can easily construct parameterizations of Hadamard matrices depending on a number of free phases at least  for a non-prime $n$. That means that the set of the moduli equations has to be  split in some sub-sets and for each such sub-set the solutions are in $\underbrace{S^1\otimes\dots\otimes S^1}_{k\,\, factors}$, where $k$ is the number of  arbitrary  phases parameterizing  the considered sub-set.  But this could be equivalent to the existence of a rational parameterization for the equations defining this sub-set. Unfortunately the best studied case  and the best results are for algebraic curves; see \cite{Ko}, Theorem 14, for a flavour of recent results. The study of surfaces, three-fold, etc. is at the beginning and  until now the theory was developed only for the simplest varieties, the so called rationally connected varieties  \cite{Ko}. From what we said before one may conclude that the parameterization of complex Hadamard matrices could be  an interesting example  of the parameterization of meromorphic varieties, which could be a mixing between a rational parameterisation and a  parameterisation of hyperelliptic curves.
 Thus the theoretical instrument  for   the  parameterization of  complex Hadamard matrices seems to  exist, the  challenging  problem being the transformation of the existing theorems into a symbolic manipulation software program able to find after a reasonable computer time  explicit solutions  at least for moderate values of $n$.

\section{Conclusion}

All the results obtained  for the complex Hadamard matrices can be used for the
construction of {\it real} Hadamard matrices the only supplementary constraint being the
natural one $n=4\,m$. We believe that the Hadamard conjecture can be solved in
our formalism since unlike the classical combinatorial approach we have also at our disposal $(n-1)(n-2)/2$ phases, and the problem is to guess the pattern of $0$ and $\pi$ taken by them. 

Conversely many constructions from the theory of real
 Hadamard matrices can be extended to the complex case. For example a complex
 conference matrix will be a matrix with $a_{ii}=0,\,\, i=1,\dots,n$ and
 $|a_{ij}|=1/\sqrt{n}$ such that 
$$W\,W^* =\frac{n-1}{n}$$
It is not difficult to construct complex conference matrices, in fact it is a
  simpler problem than the construction of complex Hadamard matrices
 because the equations $a_{ii}=0,\,\,i=2,\dots,n-1$ imply the determination of $2(n-2)$
 parameters which simplify the other equations.

We give a few examples:

\[
W_4=\frac{1}{2}\left( \begin{array}{cccc}
0&1&1&1\\
1&0&-e^{it}&e^{it}\\
1&e^{it}&0&-e^{it}\\
1&-e^{it}&e^{it}&0
\end{array}\right)
\]

and
\begin{eqnarray}
W_6=
\frac{1}{\sqrt{6}}\left(
 \begin{array}{cccccc}
0&1&1&1&1&1\\
1&0&-e^{i\alpha} &-e^{i\alpha}&e^{i\alpha} &e^{i\alpha} \\
1&-e^{i\alpha} &0&e^{i\alpha }&-e^{i(\alpha -\beta)}&e^{i(\alpha -\beta)}\\
1&-e^{i\alpha} &e^{i\alpha }&0&e^{i(\alpha -\beta)} &-e^{i(\alpha -\beta)} \\
1&e^{i\alpha} &-e^{i(\alpha+\beta) }&e^{i(\alpha+\beta)} &0&-e^{i\alpha }\\
1&e^{i\alpha} &e^{i(\alpha+\beta)} &-e^{i(\alpha+\beta)} &-e^{i\alpha }&0\\
\end{array}
\right)\nonumber
\end{eqnarray}
where the second  depends on two arbitrary phases. They are useful because if $W_n$ is a complex conference matrix then

$$M_{2 n}= \frac{1}{\sqrt{2}}\left(
\begin{array}{cc}

W_n +{{\textstyle I_n}\over\sqrt{\textstyle n}} & W_n^* -{{\textstyle I_n}\over\sqrt{\textstyle n}} \\
&\\
W_n -{{\textstyle I_n}\over\sqrt{\textstyle n}} &- W_n^* - {{\textstyle I_n}\over\sqrt{\textstyle n}}

\end{array}\right)$$
is a complex Hadamard matrix of order $2 n$.

In this paper we have used  convenient parameterisations of unitary matrices  that allowed us getting  a set of $(n-2)^2$ polynomial equations whose solutions will give all the posible parameterisations for Hadamard matrices. Unfortunately the system is very complicated and only particular solutions have been found; thus from a pragmatical point of view the most important issue would be the design of software packages for solving these equations.
\vskip3mm
 {\bf Acknowledgements.} The work was completed while the author was a visitor at the Institute for Theoretical Physics, University of Bern in the frame of the Swiss National Science Foundation Program `` Scientific Co-operation between Eastern Europe and Switzerland (SCOPES 2000-2003)''.
It is a pleasure for me  to thank Professor H. Leutwyller for many interesting discussions. Also I want 
 to thank Professor J. Gasser for the warm hospitality extended to me during my stay in Bern.

\vskip2.5cm

\end{document}